\documentclass[prx,nobalancelastpage,a4paper,notitlepage,twocolumn,superscriptaddress,10pt, longbibliography]{revtex4-1}
\usepackage{amssymb,amsfonts,amsmath,graphicx,color,mathptm,times,amssymb,hyperref}
\usepackage[centering,hmargin=2cm,vmargin=2.5cm]{geometry}
\usepackage[utf8]{inputenc}
\usepackage{soul}
\usepackage[titletoc,title]{appendix}
\usepackage{subfloat}

\newcommand{\bra}[1]{\langle#1 \vert}
\newcommand{\ket}[1]{ \vert #1\rangle}
\newcommand{\braket}[2]{\langle#1 \vert #2\rangle}

\newcommand{\avg}[1]{\langle #1 \rangle}
\newcommand{\tr}{\mbox{Tr}}
\newcommand{\re}{\text{Re}}
\newcommand{\im}{\text{Im}}

\begin{document}
\date{\today}
\title{The power of one qumode for quantum computation}

\author{Nana Liu}
\email{Nana.Liu@physics.ox.ac.uk}
\affiliation{Clarendon Laboratory, Department of Physics, University of Oxford, Oxford OX1 3PU, United Kingdom}

\author{Jayne Thompson}
\affiliation{Centre for Quantum Technologies, National University of Singapore, 3 Science Drive 2, Singapore 117543}

\author{Christian Weedbrook}
\affiliation{CipherQ Corp., Toronto, ON M5B 2G9, Canada}

\author{Seth Lloyd}
\affiliation{Department of Mechanical Engineering and Research Laboratory of Electronics, Massachusetts Institute of Technology, Cambridge MA 02139, USA}

\author{Vlatko Vedral}
\affiliation{Clarendon Laboratory, Department of Physics, University of Oxford, Oxford OX1 3PU, United Kingdom}
\affiliation{Centre for Quantum Technologies, National University of Singapore, 3 Science Drive 2, Singapore 117543}
\affiliation{Department of Physics, National University of Singapore, 3 Science Drive 2, Singapore 117543}
\affiliation{Center for Quantum Information, Institute for Interdisciplinary Information Sciences, Tsinghua University,
Beijing, 100084, China}

\author{Mile Gu}
\email{cqtmileg@nus.edu.sg}
\affiliation{School of Physical and Mathematical Sciences, Nanyang Technological University, Singapore 639673, Singapore}
\affiliation{Complexity Institute, Nanyang Technological University, 639673, Singapore}
\affiliation{Centre for Quantum Technologies, National University of Singapore, 3 Science Drive 2, 117543, Singapore}

\author{Kavan Modi}
\email{kavan.modi@monash.edu}
\affiliation{School of Physics and Astronomy, Monash University, Victoria 3800, Australia}

\begin{abstract}
Although quantum computers are capable of solving problems like factoring exponentially faster than the best-known classical algorithms, determining the resources responsible for their computational power remains unclear. An important class of problems where quantum computers possess an advantage is phase estimation, which includes applications like factoring. We introduce a new computational model based on a single squeezed state resource that can perform phase estimation, which we call the power of one qumode. This model is inspired by an interesting computational model known as deterministic quantum computing with one quantum bit (DQC1). Using the power of one qumode, we identify that the amount of squeezing is sufficient to quantify the resource requirements of different computational problems based on phase estimation. In particular, we can use the amount of squeezing to quantitatively relate the resource requirements of DQC1 and factoring. Furthermore, we can connect the squeezing to other known resources like precision, energy, qudit dimensionality and qubit number. We show the circumstances under which they can likewise be considered good resources.
\end{abstract}

\maketitle

\textit{Introduction.---} Quantum computing is a rapidly growing discipline that has attracted significant attention due to the discovery of quantum algorithms that are exponentially faster than the best-known classical ones \cite{dj, shor, grover, harrowlloyd}. One of the most notable examples is Shor’s factoring algorithm \cite{shor}, which has been a strong driver for the quantum computing revolution. However, the essential resources that empower quantum computation remain elusive. Knowing what these resources are will have both great theoretical and practical consequences. This knowledge will motivate designs that takes optimal advantage of such resources. In addition, it may further illuminate the quantum-classical boundary.

In pure state quantum computation, it is known that entanglement is a necessary resource to achieve a computational speed-up \cite{jozsalinden}. This is no longer true for mixed-state quantum computation and it is unclear if a single entity can quantify the computational resource in these models. Or if multiple resources appear as candidates, it has not been made explicit what the relationship is between these different resources. One notable example is the \emph{deterministic quantum computation with one quantum bit} (DQC1) model \cite{knill1998power}. This model contains little entanglement and purity \cite{white, datta2007}. Yet it can solve certain computational problems exponentially faster than the best-known classical algorithms by using a highly mixed target state and a single pure control qubit.  However, it is unclear how to compare the resources needed for DQC1 and factoring on an equal footing since there is currently no example of both of these two problems solved using the same model. Although suggestions have been made that factoring requires more resources than DQC1 \cite{parkerplenio}, a direct quantitative relation between the two is still lacking. 

To address this challenge, in this paper we propose a continuous-variable (CV) extension of DQC1 by replacing the pure qubit with a CV mode, or qumode.  We call this new model the \emph{power of one qumode}. We demonstrate that our model is capable of reproducing DQC1 and factoring in polynomial time. This enables us to identify a CV resource in our model, called squeezing, to compare factoring and DQC1 on the same level. Squeezed states are also useful resources in other contexts, like gaining a quantum advantage in metrology \cite{caves, monras, pinel} and in CV quantum computation \cite{lloydCV, mile2009}. 

The term `squeezing' could refer to either the squeezing parameter $r$ or the squeezing factor $s_0=\exp(r)$. For quantifying resources in the context of computational complexity, it is important to make a distinction between these two definitions since they are exponentially separated. We motivate our use of the squeezing factor over the squeezing parameter by showing how it can be interpreted as inverse precision, which is a known resource in computational complexity \cite{algorithms}.

By inputting a squeezed state as the pure qumode, we can perform both the hardest problem in DQC1 and phase estimation. We can relate the squeezing factor to the degree of precision in phase estimation and the total computation time. As an application, we can show there exists an algorithm using our model that can factor an integer efficiently in time and it requires a squeezing factor that grows exponentially with the number of bits to encode this integer. Another algorithm in our model can recover DQC1 with no squeezing.

A further way of interpreting the squeezing factor is through the dimensionality of a qudit that can be encoded in the squeezed state, which we later examine.  In some cases, the squeezing factor can also be considered as an energy resource, while the squeezing parameter can be interpreted in terms of the number of qubits. We discuss all these connections more precisely later in this paper.

Before moving on, let us remark that our architecture is an example of a hybrid computer: it jointly uses both discrete and CV systems. A similar hybrid model using a pure target state was given by Lloyd \cite{lloydhybrid} to find eigenvectors and eigenvalues.  Hybrid models for computing are interesting in their own right for providing an alternative avenue to quantum computing that bypasses some of the key obstacles to fully CV computation using linear optics or fully discrete-variable models \cite{lloydhybrid, hybridbook}. This creates an important best-of-both-worlds approach to quantum computing.
\begin{figure}[ht!]
\centering
\includegraphics[scale=0.3]{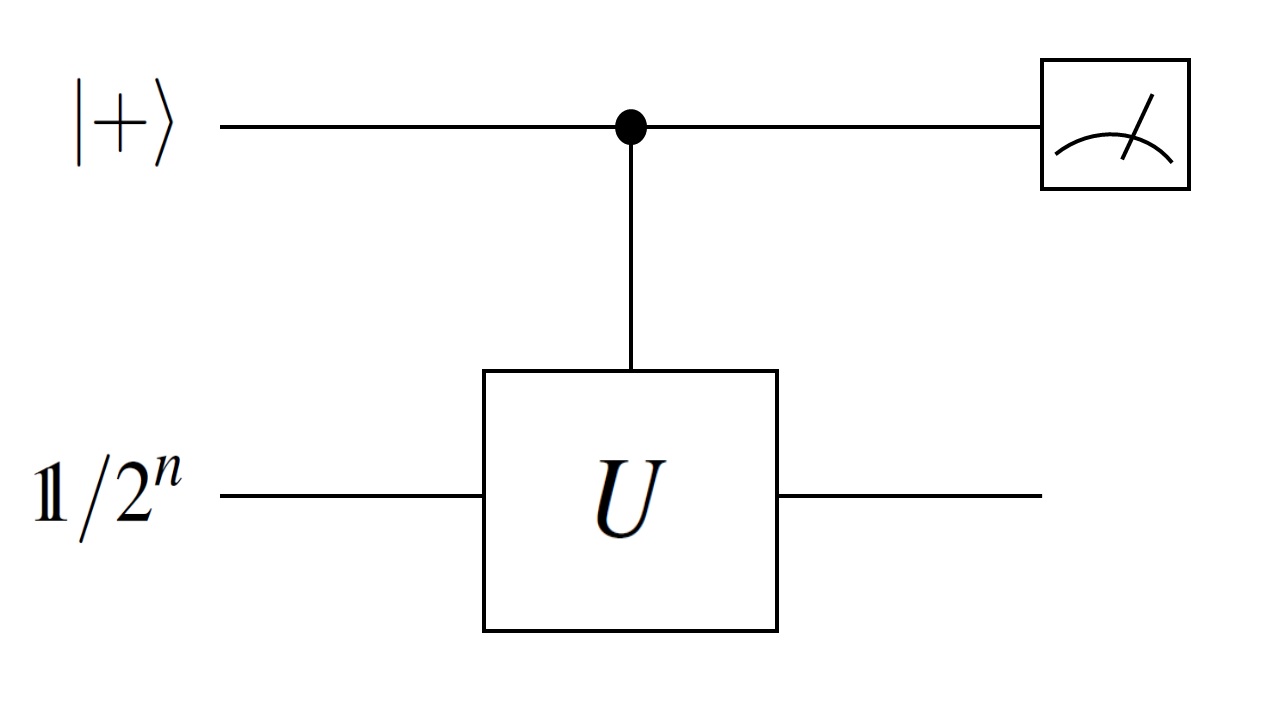}
\caption{\label{fig1}DQC1 circuit. The control state is $\ket{+}$ and the target state is $n=\log_2N$ qubits in a maximally mixed state. Here $U$ is an $N \times N$ matrix and one can measure the final average spin of the control state to recover
the normalised trace of $U$.}
\end{figure}

\begin{figure}[ht!]
\centering
\includegraphics[scale=0.3]{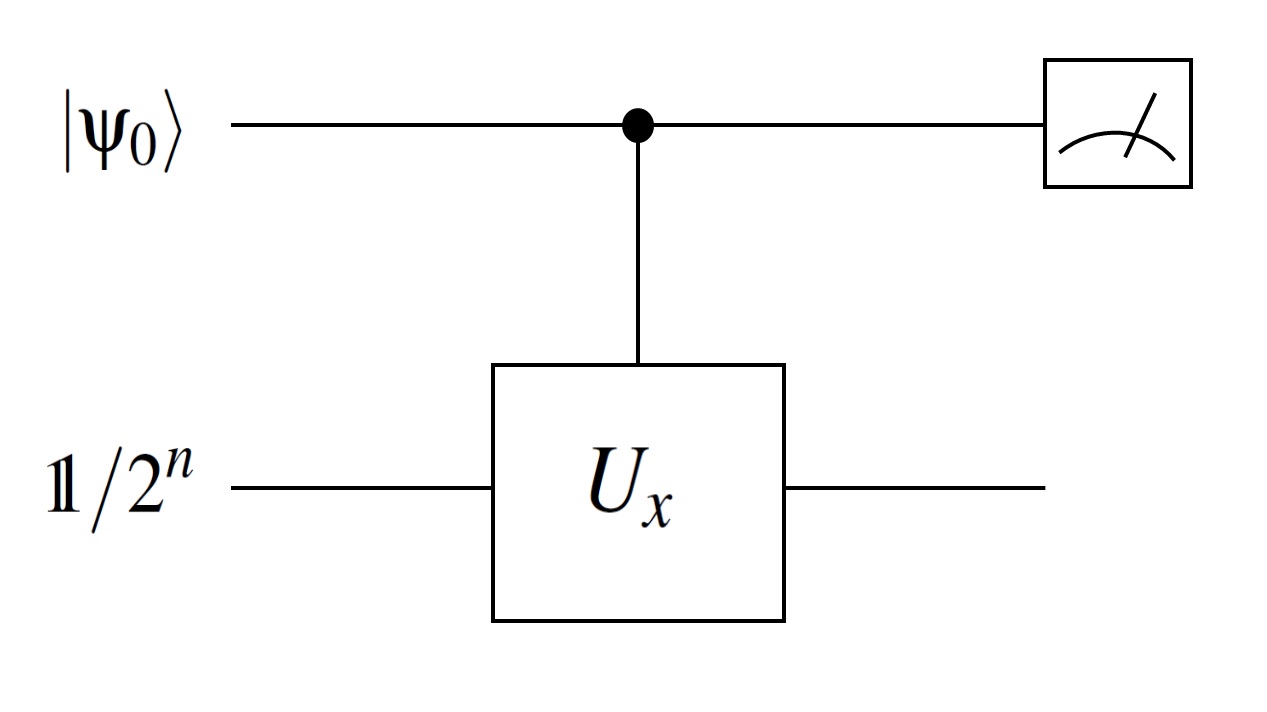}
\caption{\label{fig2}Power of one qumode circuit. We can have a squeezed state $\ket{\psi_0}$ as the control state. The target state consists of $n=\log_2 N$ qubits in a maximally mixed state as in DQC1. Here $U_x \equiv \exp(i x H \tau/x_0 )$ where $x_0$ is a constant and $\tau$ is the gate running time. Its relationship to the unitary in DQC1 is $U_x=U^{x \tau/x_0}$. We make final measurements of the control state in the momentum basis. The momentum measurements in this model can be used to recover the normalised trace of an $N \times N$ matrix $U$ and also to factor the integer $N$. }
\end{figure}

\textit{DQC1.---} The most difficult  DQC1 problem, called DQC1-complete, is estimating the normalised trace of a unitary matrix \cite{shorjordan, shepherd}. This problem turns out to be important for a diverse set of applications \cite{fidelitydecay, knill2001, shorjordan}, such as estimating the Jones polynomial. Computing the normalised trace of a unitary begins with a pure control qubit in the state $\ket{+}=(\ket{0}+\ket{1})/\sqrt{2}$ and a target register made up of $n$ qubits that are in a fully-mixed state $\openone/2^n$. Next, the control and target registers interact via a controlled-unitary operation, represented by $\Gamma_U = \ket{0} \!\bra
{0} \otimes$ $\openone$ $+\ket{1}\!\bra{1} \otimes U$ where $U$ acts on the qubits in the target register. The control qubit measurement statistics yields the normalised trace of $U$, i.e.,  $\avg{\sigma_x+i \sigma_y} = \tr(U) / 2^n$. The circuit for DQC1 is shown in Fig.~\ref{fig1}. To estimate the normalised trace to within error $\delta$, that is, $\tr(U)/2^n \pm \delta$, we need to run the computation $T_{\text{DQC1}} \sim1/[\text{min}\{\text{Re}(\delta), \text{Im}(\delta)\}]^2$ times \cite{datta2005}. Since $\delta$ is independent of the size of $U$, this computation is efficient and DQC1 has an exponential advantage over the best-known classical algorithms \cite{animeshphd}.

\textit{One qumode model.--} In this paper we extend DQC1 by replacing the pure control qubit with a pure CV state (qumode), while keeping the target register the same. The total input state in our model is thus a hybrid state of discrete-variables and a CV. See Fig.~\ref{fig2} for the circuit diagram of our model. We first show how our model can perform the quantum phase estimation algorithm \cite{cleve}. We use this to efficiently compute (in time) a DQC1-complete problem, thus showing that this model contains DQC1. Next, we show that our model can perform Shor's factoring algorithm, which is based on the phase estimation algorithm. 

The aim in the phase estimation problem is to find the eigenvalues of a Hamiltonian, $H \ket{u_j}=\phi_j \ket{u_j}$. The complete set of eigenvalues of $H$ is given by $\{ \phi_j \}$. We encode the Hamiltonian $H$ into a unitary transformation, $C_U$, that acts on the hybrid input state. We call $C_U$ the hybrid control gate and is defined as $C_U=\exp(i \, \hat{x} \otimes H \tau/x_0)$, where the position operator $\hat{x}$ acts on the qumode \footnote{It is also possible to define a control gate controlled on the particle number operator instead of $\hat{x}$. However, analytical solutions in this case are not straightforward and for our purposes it suffices to look at our current hybrid control gate.} and $\tau$ is the running time of the hybrid gate. Here $x_0 \equiv 1/\sqrt{m \omega}$, where $m, \omega$ are the mass and frequencies of the harmonic oscillator corresponding to the qumode \footnote{Here we use natural units $\hbar=1=c$.}. Like the control gate $\Gamma_U$ in DQC1, the hybrid control gate can also be decomposed into elementary operations (see Appendix A). If the qumode is in a position eigenstate $\ket{x}$ and $\ket{u_j}$ is a state of target register qubits, the action of the hybrid control gate is
\begin{gather}
C_U \ket{x}\otimes \ket{u_j}=\ket{x} \otimes U_x \ket{u_j}
= \ket{x} \otimes e^{i \phi_j x \tau/x_0} \ket{u_j},
\end{gather}
where $x$ is the eigenvalue of $\hat{x}$ and $U_x \equiv \exp(i x H \tau/x_0 )$. In our model, we apply $C_U$ to a maximally mixed state of $n$ qubits and a qumode state $\ket{\psi_0} = \int G(x) \ket{x} dx$. $G(x)$ is the wave-function of the initial qumode in the position basis. After implementing this gate, the target register is discarded, and the qumode is in the state
\begin{gather}
\rho_f=\frac{1}{2^n} \iint G(x) G^*(x') \tr[e^{i(x-x')H \tau/x_0}] \ket{x}\!\bra{x'} \text{d}x \, \text{d}x'.
\end{gather}

Next, we measure this state in the basis of the momentum operator $\hat{p}$ \footnote{Operators $\hat{x}$ and $\hat{p}$ satisfy the canonical commutator relation $[\hat{x}, \hat{p}]=i$}, i.e., $\braket{p} {\rho_f \vert p}$. This measurement yields the momentum probability distribution 
\begin{align} \label{eq:momentum1}
\mathcal{P}(p) &=\frac{1}{2^n} \sum_m \iint G(x)G^*(x') e^{i(x-x')\phi_m \tau/x_0} \braket{p}{x}\! \braket{x'}{p} \text{d}x \, \text{d}x' \nonumber \\
        &=\frac{1}{2^n} \sum_m \mathcal{G}(\phi_m \tau/x_0-p) \mathcal{G}^*(p-\phi_m \tau/x_0),
\end{align}
where we used $\braket{p}{x}=(1/\sqrt{2\pi})\exp(-ixp)$ and the Fourier transform of $G(x)$ is denoted by $\mathcal{G}(p)=(1/\sqrt{2\pi})\int^{\infty}_{-\infty} \exp(ixp)G(x)\text{d}x$. 

If we choose our wavefunction $G(x)$ carefully, we can employ our model to recover the eigenvalues of $H$. Suppose we initialized the control mode in a coherent state $\ket{\alpha}$, chosen for its experimental accessibility \cite{knightgerry}. If we measure the probability distribution of $p_{\text{E}} \equiv px_0/\tau$ where $x_0$ and $\tau$ are known inputs and $p_{\text{E}}$ has dimensions of energy, we find (see Appendix B for a derivation)
\begin{gather}\label{eq:coherentp}
\mathcal{P}(p_{\text{E}})=\frac{\tau}{\sqrt{\pi} 2^n} \sum_{m=1}^{2^n} e^{-\tau^2 \left[p_{\text{E}}-\left(\phi_m+\frac{\text{Im}(\alpha)}{\tau}\right)\right]^2},
\end{gather}
where $\text{Im}(\alpha)$ is the imaginary component of $\alpha$ \footnote{This is equivalent to the initial expectation value of momentum of the coherent state }. We can see that the probability distribution is a sum of Gaussian distributions. It has individual peaks centred at each shifted eigenvalue $\phi_j+\text{Im}(\alpha)$ with an individual spread given by the inverse of $\tau$. By sampling this probability distribution we can infer the position of the peaks to any finite precision. Thus it is possible to perform phase estimation to arbitrary accuracy just by increasing $\tau$ alone. However, to estimate eigenvalues to a precision better than a polynomial in $n=\log_2 N$, we require $\tau$ greater than polynomial in $n=\log_2 N$. Thus the coherent state no longer suffices for Shor’s factoring algorithm, which requires high precision phase estimation. In such cases, we require a further resource that we identify to be the squeezing factor. 

A finite squeezed state is defined by $G(x) = (1 / (\sqrt{s}\pi^{\frac{1}{4}})) \text{exp}(-x^2/(2s^2))$ where $s\equiv s_0 x_0$ and $s_0$ parameterises the amount of squeezing in the momentum direction \footnote{Here $s_0$ is a real number in the range $s_0 \in [1, \infty)$.}. We call $s_0$ the squeezing factor. It's wavefunction in $x$ has a Gaussian profile with standard deviation $1/s_0$. By inputting a squeezed state into our model, the probability distribution in $p_{\text{E}}$ becomes
\begin{gather} \label{eq:pp}
\mathcal{P}(p_{\text{E}})=\frac{s_0 \tau}{2^n \sqrt{\pi}} \sum_{m=1}^{2^n} e^{-(s_0 \tau)^2(p_{\text{E}}-\phi_m)^2}.
\end{gather}
Comparing this to Eq.~\eqref{eq:coherentp} we see the coherent state plays the same role as an unsqueezed state (i.e., $s_0=1$). The method for retrieving the eigenvalues is now identical to that of the coherent state, except now we can take advantage of a large squeezing factor instead of non-polynomial gate running time. 

We can see the relationship between the squeezing factor and gate running time more explicitly. Let $T_{\text{bound}}$ be the upper bound to the total number of momentum measurements we are willing to make for phase estimation. If we need to recover any eigenvalue of the Hamiltonian to accuracy $\Delta_{\text{E}}$, the following time-energy condition is satisfied (see Appendix C for a derivation)
\begin{gather}\label{eq:tbound}
T_{\text{bound}}\tau s_0 \Delta_{\text{E}} \gtrsim 1,
\end{gather}
where $\Delta_{\text{E}}$ can be a function of the size of the Hamiltonian. In an efficient protocol the maximum total gate running time $T_{\text{bound}} \tau$ is bounded by a polynomial in $n$. When the inverse of $\Delta_{\text{E}}$ is also a polynomial in $n$, efficient phase estimation is still possible for a squeezing factor polynomial in $n$. For example, this is useful for the verification of problems in the Quantum-Merlin-Arthur (QMA) complexity class, which includes the local Hamiltonian problem \cite{localhamiltonian}. For an exponentially greater precision in phase estimation, however, an exponentially higher squeezing factor is needed. We see from Eq.~\eqref{eq:tbound} that the squeezing factor serves as a rescaling of the energy `uncertainty' $\Delta_{\text{E}}$. Similarly to phase estimation, increased squeezing can also retrieve the corresponding eigenvectors to greater precision \footnote{See Appendix D. Also see \cite{abramslloyd} for another algorithm on eigenvector retrieval}. 

We can see the precise relationship between the squeezing factor and inverse precision from Eq.~\eqref{eq:tbound} by considering when the maximum total gate running time resource is constrained. When the time resource is constant, the minimum squeezing factor required for efficient phase estimation is the inverse precision, i.e., $s_0 \sim 1/\Delta_E$. 

This relationship can be seen more intuitively by considering a problem whose solution is given by the central position $x_0$ of a squeezed state with squeezing factor $s_0$. From the central limit theorem, it requires $t \sim 1/(s_0^2 \eta^2)$ measurements of the position $x$ to get within precision $\eta=|x-x_0|$ of the centre. Thus for a fixed number of measurements (or time), the squeezing factor scales as the inverse of precision $s_0 \sim 1/\eta$. 

Another way we can see $s_0$ as the inverse precision is to consider when we are trying to resolve the distance between two adjacent Gaussian peaks $\Delta \phi$. We see later that factoring in our model is essentially this problem with $\Delta \phi \sim 1/N=1/2^n$, where $N$ is the number to be factored. Each Gaussian has standard deviation $1/s_0$. If the distance between these peaks is closer than this length scale, it becomes difficult to resolve the two peaks. Thus $1/s_0$ is the maximum resolution for $\Delta \phi$, which is another precision scale. This fact is used when we later examine the qubit and qudit encoding in our model.

\textit{Recovering DQC1.---}
We begin with an observation that the average of $\exp(ip_{\text{E}})$ can reproduce the normalised trace of $U \equiv \exp(iH)$ in the following way
\begin{gather}
\int e^{ip_{\text{E}}} \mathcal{P}(p_{\text{E}}) \text{d}p_{\text{E}}=e^{-\frac{1}{4 s_0^2}} \frac{\tr(U_{\tau})}{2^n},
\end{gather}
where $\mathcal{P}(p_{\text{E}})$ is given by Eq.~\eqref{eq:pp} and $U_{\tau}\equiv \exp(iH \tau)$. For an $N \times N$ matrix $U_{\tau}$, we use $n=\log_2 N$. If we wish to recover the normalised trace of $U$ to within an error $\delta$ (i.e., $\tr(U)/2^n \pm \delta$), we require $\tau=1$ and $T_{\text{DQC1}}$ measurements of momentum \footnote{Note that the number of momentum measurements and $p_{\text{E}}$ measurements needed are equivalent.} in our model. This is equivalent to running our hybrid gate once per momentum measurement and then averaging the corresponding values $\{ \exp(ip_{\text{E}}) \}$. 

This computation of the normalised trace is as efficient as DQC1 if $T_{\text{DQC1}}$ is independent of $N=2^n$. By employing the central limit theorem we find (see Appendix E for a derivation) 
\begin{gather}\label{eq:tF}
T_{\text{DQC1}} \lesssim \frac{F(s_0)}{[\text{min}\{\text{Re}(\delta), \text{Im}(\delta)\}]^2},
\end{gather}
where $F(s_0)=\text{sinh}(1/(2s_0^2))+\exp(-1/(2 s_0^2))$ and $F(s_0) \rightarrow 1$ very quickly with increasing $s_0$ \footnote{The $F(s_0)$ overhead is analogous to the case in DQC1 when using a slightly mixed state probe state instead of the pure state {$\ket{+}\!\bra{+}$} \cite{animeshphd}. The degree of mixedness does not affect the result that the computation is efficient. The amount of squeezing in our model thus corresponds to the degree of mixedness in the input state of DQC1. Higher squeezing corresponds to greater purity.}. Eq.~\eqref{eq:tF} shows $T_{\text{DQC1}}$ is upper bounded by a quantity dependent only on the squeezing and not on the size of the matrix. In fact, even when $s_0=1$ (equivalent to a coherent state input) our qumode model is sufficient to efficiently compute (in time) the normalised trace of $U$, thus reproducing DQC1. This can also be viewed as a consequence of $\Delta_{\text{E}}$ being independent of $N=2^n$ in Eq.~\eqref{eq:tbound}. 

\textit{Factoring using power of one qumode.---}
Factoring is the problem of finding a non-trivial multiplicative factor of an integer $N$. The classically hard part can be reduced to a phase estimation problem, where the quantum advantage in phase estimation can be exploited. We show how the corresponding phase estimation problem can be solved in our model and how much squeezing resource is required. 

Factoring can be reduced to phase estimation in the following way. There is a known classically efficient algorithm that can find a non-trivial factor of $N$ once it is given a random integer $q$ in the range $1 < q < N$ \cite{shor}. However, this algorithm relies on prior knowledge of the order $r$ of $q$, where $r$ is an integer $r \leq N$ satisfying $q^r \equiv 1 \mod N$. Thus the main difficulty lies in finding this order $r$, which is believed to be a classically hard problem. It turns out this order can be encoded into the eigenvalues of a suitably chosen Hamiltonian $H_q$. 

Here we begin with a squeezed control state and a target state of $n=\log_2 N$ qubits in a maximally mixed state. Let our hybrid control gate be $C_{U_q}=\exp(i \hat{x} \otimes H_q \tau/x_0)$. Next we choose a suitable Hamiltonian $H_q$ whose eigenvalues contain the order $r$. We define a unitary $\exp(iH_q)$ which acts on a qubit state $\ket{l \text{mod} N}$ like $\exp(iH_q)\ket{l \text{mod} N}=\ket{lq \text{mod} N}$, where $l$ is an integer $0 \leq l<N$. When $l=q^k$ for an integer $k \leq r$, $\exp(iH_q r)\ket{q^k \mod N}=\ket{q^k q^r \mod N}=\ket{q^k \mod N}$. Here the eigenvalues of $H_q$ are $2 \pi m/r$ where $m$ is an integer $1 \leq m<r$. However, for qubits in a mixed state  we have $l \neq q^k$ in general.  In these cases, we define a more general ``order'' $r_d$ where $\exp(iH_q r_d)\ket{l_d \mod N}=\ket{l_d q^{r_d} \mod N}=\ket{l_d \mod N}$. Here $r_d$ is an integer $r_d \leq r$ that divides $r$ \cite{parkerplenio} and satisfies $l_d q^{r_d} \mod N=l_d \mod N$. The integer $d$ labels the set of states $\{ \ket{l_dq^h \mod N}\}$ where $h\leq r_d$ is an integer. Thus for general $l_d$, the eigenvalues of $H_q$ can be written as $2 \pi m_d/r_d$ where $m_d$ is an integer $1 \leq m_d <r_d$. 

These eigenvalues do not give $r$ directly. However, we can always rewrite $m_d/r_d$ in the form $m/r$ since $r_d$ is a factor of $r$. In general, there will be a single fraction $m/r$ corresponding to many possible $m_d$ and $r_d$. If we call this multiplicity $c_m$ for a given $m/r$, then following Eq.~\eqref{eq:pp} we can write the $p_{\text{E}}$ probability distribution as measured by the final control state as
\begin{align}\label{eq:PDFfactor}
\mathcal{P}(p_{\text{E}}) &=\frac{s_0 \tau}{\sqrt{\pi} 2^n} \sum_{d} \sum_{m_d=0}^{r_d-1} e^{-(2 \pi s_0 \tau)^2 \left(\frac{p_{\text{E}}}{2 \pi}-\frac{m_d}{r_d}\right)^2} \nonumber \\
       &=\frac{s_0 \tau}{\sqrt{\pi} 2^n} \sum_{m=0}^{r-1} c_m e^{-(2 \pi s_0 \tau)^2 \left(\frac{p_{\text{E}}}{2 \pi}-\frac{m}{r}\right)^2}.
\end{align}
This probability distribution is a sum of Gaussian functions with amplitudes $c_m$ and centered on $m/r$. To recover the order $r$ from the above probability distribution, it is sufficient to satisfy two conditions. The first condition is to be able to recover the fractions $m/r$ to within the interval $[m/r-1/(2N^2), m/r+1/(2N^2)]$ \footnote{This ensures that $m/r$ is recovered exactly by using the continued fractions algorithm. See \cite{nandc} for an explicit demonstration}. Thus the larger the number we wish to factor, the more squeezing we need to improve the precision of the phase estimation. The second requirement is for $m$ and $r$ to be coprime, which enables us to find $r$. This requirement is satisfied with probability less than $\mathcal{O}(\ln[\ln(N)])$. 

Subject to the above two conditions, we can compute the probability that a correct $r$ is found using the momentum probability distribution in our model. We derive in Appendix F the number of runs $T_{\text{factor}}<\mathcal{O}(\ln[\ln(N)])/\text{erf}(\pi s_0 \tau/N^2)$ needed to factor $N$, which is inversely related to the probability of finding a correct $r$. In the large $N$ limit, to achieve the same efficiency as Shor's algorithm using qubits, which is $T_{\text{factor}} \sim \mathcal{O}(\ln[\ln(N)])=\mathcal{O}(\ln[\ln(N)]) T_{\text{bound}}$ \footnote{Note that $T_{\text{bound}}$ in this case corresponds to the number of momentum measurements needed to find the correct eigenvalue of the Hamiltonian. From the eigenvalue, one still needs an extra classically efficient step to find the factor, so $T_{\text{factor}}>T_{\text{bound}}$.} it is thus sufficient to choose 
\begin{gather} \label{eq:ssufficient}
s_0 \tau \sim 2^{2n}.
\end{gather}
This can also be derived from Eq.~\eqref{eq:tbound} using $\Delta_{\text{E}}=2\pi/(2N^2)$, where $T_{\text{bound}} \sim 1$. If we let $s_0=1$ for the coherent state, this requires total computing time to scale exponentially with the size of the problem (i.e., $\log_2 N$). Thus to ensure polynomial total computing time, we can choose instead $\tau \sim 1$ and $s_0 \sim 2^{2n}$. 

\textit{Comparing squeezing to other resources.---}
We saw that the squeezing factor can be interpreted as an inverse precision since the two quantities are also polynomially related. There are also other quantities polynomially related to the squeezing factor like energy and the dimensionality of the qubit that can be encoded in our squeezed state. We discuss their relationship to the squeezing factor and in what ways they can and cannot also be considered resources.

\textit{Energy}. Energy may be considered a resource if it is required in the initial preparation of the necessary input states. In a quantum optical setting, for example, energy is required for preparing a squeezed state resource. The minimum energy $E_{\text{min}}$ required is that needed to create the number of particle excitations $\avg{n_p}$ corresponding to a certain amount of squeezing since $E_{\text{min}} \propto \avg{n_p}$. The number of particle excitations is itself regularly considered as the primary resource in the context of quantum metrology. For our squeezed state $\langle n_p \rangle =\text{sinh}^2(\ln (s_0))$, where for a large squeezing factor $\avg{n_p} \propto s_0^2$. Thus energy and the squeezing factor are polynomially related.

This interpretation of the squeezing factor as an energy can help us understand why $s_0$ of the order $\mathcal{O}(\exp(n))$ is necessary for factoring in our algorithm. We can consider performing factoring in our model as swapping $m=\log_2 N$ pure control qubits in the qubit factoring protocol with a single qumode. A simple example to illustrate this phenomenon is to consider a simple computation $\ket{0}^{\otimes \mu} \rightarrow \ket{1}^{\otimes \mu}$. Suppose the computation is performed using $\mu$ qubits encoded in $\mu$ two-level atoms. Let the energy gap between the ground ($\ket{0}$) and the first excited state ($\ket{1}$) be $\Delta E$. Then a total energy of $\mu \Delta E$ is required for the computation. If we use a single CV mode instead, for instance, a harmonic oscillator with $2^{\mu}$ energy levels, the total energy required to perform this computation is $2^{\mu} \Delta E$, which has exponential scaling in $\mu$. This is very similar to the exponential scaling in $\log_2 N$ observed in our model.

However, there are also two reasons why it is not ideal to consider energy as a resource.  Firstly, having no energy does not guarantee that the computational power of a high squeezing factor cannot be achieved. An example is spin-squeezing in the case of energy-degenerate spin states. Secondly, having large amounts of available energy also does not guarantee more efficient computation. If we instead use a coherent state with high coherence $\alpha$ and hence large energy (since $\langle n_p \rangle=|\alpha|^2$), we still cannot factor in polynomial time.

\textit{Qudit dimensionality.} The GKP (Gottesman-Kitaev-Preskill) encoding \cite{GKP} allows one to encode a qudit, or a discrete variable quantum state with $D$ dimensions \footnote{D=2 is equivalent to a qubit.} into a CV mode. We use this encoding scheme as an illustration. This can work for CV states whose probability distribution (in momentum, for example) can be described as a sum of Gaussian functions each with standard deviation $w$ and neighbouring centers are separated by a distance $\Delta \phi$. Since the precision associated with each peak is on the order $w$, we can fit a total of $\Delta \phi/w$ distinguishable copies of this distribution where each copy is separated from its neighbour by a unit $\Delta \phi/w$ along the momentum axis. If we represent each degree of freedom by one such distribution, then there are $D=\Delta \phi/w$ degrees of freedom available to this CV state just by displacement in momentum. These $D$ degrees of freedom can be mapped onto a qudit of dimensionality $D$.

Given an encoding like GKP, we can write $D \sim s_0 \Delta \phi$ since in our case $w=1/s_0$. Thus here $s_0$ is interpreted as the inverse precision $1/w$. Since $\Delta \phi$ is the distance between adjacent Gaussian peaks in our probability distribution $P(p_E)$, to accomplish factoring, we require $s_0=2^{2n}=N^2$ and $\Delta \phi=1/N$, so $D=N$. For DQC1, $s_0=1$ and $D=2$ (since we only need a single qubit). Thus $D$ and $s_0$ are also polynomially related.

\textit{Qubit number.} A qudit of dimension $D$ is equivalent to $m=\log_2 D$ pure qubits, where $D$ is polynomially related to $s_0$. Thus for factoring, the required number of control qubits in our algorithm scales as $m \sim \mathcal{O}(\text{poly}(n))$ compared to $m=1$ for DQC1, where $n=\log_2 N$ is the number of target register qubits. Here we see that the number of qubits for the two problems are not exponentially separated. There is an  important result of Shor and Jordan \cite{shorjordan}, which compares the computational power of DQC1 with an $n$-qubit target register and a model that is an $m$-control qubit extension of DQC1. Their result claims that if $m$ is logarithmically related to $n$, then this model still has the same computational power as DQC1. On the other hand, if $m$ is polynomially related to $n$, then this model is computationally harder than DQC1. If we use $n=\log_2 N$, then the Shor and Jordan result make clear that the number of control pure qubits $m$ in these two different models are not separated exponentially, even though one model has higher computational power. However, like the time resource in these two models, $D=2^m$ in these two models are exponentially separated, which suggests that $D$ may be preferred over $m$, in the context of these particular algorithms, as a good quantifier for a computational resource. 

That the required number of control qubits scales as $m \sim \mathcal{O}(\text{poly}(n))$ is not too surprising since we observe a similarity between our model and standard phase estimation. Our model has more in common with standard phase estimation than DQC1, even though it is a hybrid extension of DQC1. We can see that by taking the average of momentum measurements in our model, we obtain the average of the eigenvalues of the Hamiltonian. The momentum average, however, does not give the normalised trace of the unitary matrix $U$ as may be expected from DQC1. This can be understood by taking a discretized version of our model, where one uses instead $\ket{x}$ for $x=0, 1, 2, ...,N$. Then the circuit reduces to the standard phase estimation circuit, which requires the $m=\log_2 N$ pure control qubits which we traded for a single qumode. From this, we can also see that our model using an infinite squeezing factor is an analogue of the standard phase estimation using an infinite number of qubits, which in both models allow us to attain infinite precision in phase estimation.

We add that this comparison with standard phase estimation further strengthens our claim that $s_0 \sim 2^{2n}=N^2$ is sufficient and maybe even necessary for factoring the number $N$. Suppose if we instead only need an exponentially smaller squeezing factor for factoring in a new algorithm. This would imply that a new algorithm performed on the qubit phase estimation circuit (i.e., the qubit analogue of our algorithm) exists that can solve factoring with exponentially fewer control qubits compared to the currently known qubit phase estimation algorithm.

While qumodes like squeezed states can be used as a way of encoding qudits and qubits \cite{GKP, terhal, schoelkopf}, the squeezing factor is still a resource that should be considered in its own right. Its emphasis over qudits is important for practical considerations. The practical advantages of considering qumode resources, in general, are that CVs typically use affordable off-the-shelf components and widely leveraged quantum optics techniques. They also have higher detection efficiencies at room temperature and can be fully integrated into current fiber optics networks \cite{braunstein2005, christian2012}.

\textit{Summary.---}
A computation is a physical process and the amount of available physical resources can limit the power of a computation.  In the power of one qumode model, we demonstrate how the squeezing factor can be viewed as a resource to quantitatively compare the difficulty of phase estimation problems like factoring and the hardest problem in the DQC1 computational class. Our model thus provides a unifying framework in which to compare the resources required for both DQC1 and factoring as well as other problems based on phase estimation. In addition, we also explore the trade-off relations between the squeezing factor, the running time of the computation and the interaction strength in our model.

The physical resources commonly discussed as computational resources are time, space and inverse precision. The definitions of computational complexity classes are also based on these \cite{gareycomputers, cook, algorithms}. We identify that squeezing can also be interpreted in terms of one of these resources: inverse precision. Furthermore, we can relate the squeezing factor to energy and qudit dimensionality. This highlights very explicitly the different ways one can quantify computational power.
\bibliography{151016arxiv}
\appendix
\section{Reducing $C_U$ gate to elementary operations}
We note that in DQC1, there is a method of reducing the control gate $\Gamma_U=\ket{0}\bra{0} \otimes \openone+\ket{1}\bra{1} \otimes U$ in terms of elementary (e.g. one or two qubit) circuits \cite{animeshphd}. The analogous gate in the power of one qumode model is the hybrid control gate $C_U=\exp(i \hat{x} \otimes H \tau/x_0)$, where we now set $\tau=x_0$ for convenience. We demonstrate how this gate can also be reduced to elementary operations to further clarify the relationship between DQC1 and the power of one qumode model.

We first write down the DQC1 set-up. The DQC1 set-up begins with a polynomial sequence of elementary (e.g. one or two qubit) gates $\{u_k=\exp(ih_k)\}$. We define the product of these gates to be $\prod_k u_k \equiv U=\exp(iH)$. The next step is to implement a control-unitary on each $u_k$, so our collection of elementary gates is now transformed into the set  $\{\lambda_u \equiv \ket{0}\bra{0} \otimes \openone+\ket{1}\bra{1} \otimes u_k\}$. The product of these gates will recover the controlled-unitary operation $\Gamma_U=\ket{0}\bra{0}\otimes \openone+\ket{1}\bra{1}\otimes U$ appearing in the description of DQC1, since
\begin{align}
\prod_k \lambda_u &=\prod_k \ket{0}\bra{0} \otimes \openone+\ket{1}\bra{1} \otimes u_k \nonumber\\
&=\ket{0}\bra{0}\otimes \openone+\ket{1}\bra{1} \otimes \prod_k u_k \nonumber \\
&=\ket{0}\bra{0}\otimes \openone+\ket{1}\bra{1}\otimes U=\Gamma_U.
\end{align}
The analogus requirement for the power of one qumode model is to begin from a polynomial sequence of elementary gates which can form the hybrid control-unitary operation $C_U=\exp(i\hat{x}\otimes H)$. We show how this can be achieved.

Let us begin with the same set of elementary gates $u_k=\{\exp(ixh_k)\}$. Instead of implementing the usual control-unitary on each $u_k$, we implement a \textit{hybrid} control unitary on each $u_k$. This means our set of elementary gates is modified into the new set $\{c_u \equiv \exp(i \hat{x} \otimes h_k)\}$. We can take the product of these operations and recover $C_U$ in the following way
\begin{align}
\prod_k c_u &=\prod_k \exp(i \hat{x} \otimes h_k)=\prod_k \int dx \ket{x}\bra{x} \otimes e^{ixh_k} \nonumber \\
&=\int dx \ket{x}\bra{x} \otimes \prod_k e^{ixh_k} \nonumber \\
&=\int dx \ket{x}\bra{x} \otimes e^{ixH}=e^{i\hat{x}\otimes H}=C_U,
\end{align}
where $x$ is a number and we used 
\begin{align}
\prod_k e^{ixh_k}\equiv e^{ixH},
\end{align}
which must be satisfied for all $x$. This condition, combined with the definition that $\prod_k u_k=\prod_k \exp(ih_k)=\exp(iH)=U$, implies that $[h_k, h_{k'}]=0$ for all $k,k'$ in the product $\prod_k$ \footnote{This also implies $H=\sum_k h_k$.}. Equivalently, this means $\{u_k\}$ must be a commuting set of operators.

We can show that such a set $\{u_k\}$ where $U=\exp(iH)=\prod_k u_k$ exists for the factoring problem. We know that factoring the number $N$ is equivalent to finding the order $r$ of a random integer $q$ where $1<q<N$, which requires $U \ket{1 \mod N}=\exp(iH) \ket{1 \mod N}=\ket{q \mod N}$. Since $q$ is an integer, we can make a binary decomposition $q-1=2^0b_0+2^1b_1+2^2b_2+...+2^{f}$ where $f$ is an integer and $b_j=0,1$. Then if choose $u_k$ to be an elementary operation defined by $u_k \ket{1 \mod N}=\ket{(1+2^kb_k) \mod N}$, we can see that all operators in $\{u_k\}$ commute and $\prod_{k=0}^f u_k \ket{1 \mod N}=\ket{q \mod N}=U \ket{1 \mod N}$. 
\section{Coherent state in power of one qumode}
Suppose we begin with a coherent state $\ket{\alpha}$ in our model. The coherent state can be written in the position basis as
\begin{gather}
\ket{\alpha}=\int \braket{x}{\alpha} \ket{x} dx,
\end{gather}
whose position wavefunction is  
\begin{gather}
\braket{x}{\alpha}=\left(\frac{1}{\pi x_0^2}\right)^{\frac{1}{4}} e^{-\frac{1}{2x_0^2}(x-\text{Re}(\alpha))^2}e^{i\text{Im}(\alpha)x/x_0}e^{-\frac{i}{2}\text{Re}(\alpha)\text{Im}(\alpha)},
\end{gather}
where $x_0\equiv 1/\sqrt{m \omega}$ and $m, \omega$ are the mass and frequency scales of the corresponding quantum harmonic oscillator. 

By using $G(x) \equiv \braket{x}{\alpha}$ in Eq.~\eqref{eq:momentum1}, we find the momentum probability distribution of the final control state to be
\begin{align}
\mathcal{P}(p) &=\frac{1}{2^n} \sum_m \iint G(x)G^*(x') e^{i(x-x')\phi_m \tau/x_0} \braket{p}{x}\! \braket{x'}{p} \text{d}x \, \text{d}x' \nonumber \\
        &=\frac{x_0}{\sqrt{\pi} 2^n} \sum_{m=0}^{2^n} e^{-x_0^2 \left[p-\frac{\tau}{x_0}\left(\phi_m+\frac{\text{Im}(\alpha)}{\tau}\right)\right]^2}.
\end{align}
If we measure variable $p_{\text{E}} \equiv p x_0/\tau$ (where inputs $x_0$ and $\tau$ are initially known), the probability distribution for $p_{\text{E}}$ is
\begin{gather}
\mathcal{P}(p_{\text{E}})=\frac{\tau}{\sqrt{\pi} 2^n} \sum_{m=0}^{2^n} e^{-\tau^2 \left[p_{\text{E}}-\left(\phi_m+\frac{\text{Im}(\alpha)}{\tau}\right)\right]^2}.
\end{gather}
Thus the coherent state can be used for phase estimation, where the accuracy of the phase estimation improves with increasing running time of the hybrid gate. 
\section{Phase estimation with power of one qumode}
Suppose we want to recover any eigenvalue of our Hamiltonian to accuracy $\Delta_{\text{E}}$. The total number of $p_{\text{E}}$ measurements required for an average of one success is
\begin{gather} \label{eq:tprob1}
T_{\text{measure}} \sim \frac{1}{P_{\Delta_{\text{E}}}},
\end{gather}
where $P_{\Delta_{\text{E}}}$ is the probability of retrieving the eigenvalues to within the interval $[\phi_j-\Delta_{\text{E}}, \phi_j+\Delta_{\text{E}}]$. Using Eq.~\eqref{eq:pp} we find
\begin{align}
P_{\Delta_{\text{E}}}&\equiv \mathcal{P} \left(p_{\text{E}};\left|p_{\text{E}}-\phi_n\right| \leq \Delta_{\text{E}}\right) \nonumber \\
                   & =\frac{s_0 \tau}{\sqrt{\pi}2^n} \sum_{l=1}^{2^n} \int^{\phi_l+\Delta_{\text{E}}}_{\phi_l-\Delta_{\text{E}}} \sum_{m=1}^{2^n}e^{-(s_0 \tau)^2 \left(p_{\text{E}}-\phi_m\right)^2} \, \text{d}p_{\text{E}} \nonumber \\
                  & \equiv P(l=m)+P(l \neq m),
\end{align}
where
\begin{align}
P(l=m) &=\frac{s_0 \tau}{\sqrt{\pi}2^n} \sum_{m=1}^{2^n} \int^{\phi_m+\Delta_{\text{E}}}_{\phi_m-\Delta_{\text{E}}} e^{-(s_0 \tau)^2 (p_{\text{E}}-\phi_m)^2}\text{d}p_{\text{E}} \nonumber \\
                  &=\text{erf} \left(s_0 \tau \Delta_{\text{E}}\right)
\end{align}
and $P(l \neq m)=(1/2^n)\sum_{l \neq m=1}^{2^n}\{\text{erf}\{s_0 \tau [(\phi_l-\phi_m)/r+\Delta_{\text{E}}]\}-\text{erf}\{s_0 \tau [(\phi_l-\phi_m)/r-\Delta_{\text{E}}]\}\}>0$. 
These two contributions to the total probability distribution $P_{\Delta_{\text{E}}}$ can be interpreted in the following way. $P(l=m)$ is the probability of finding $\phi_n$ to within $\Delta_{\text{E}}$ 
if the Gaussian peaks are very far apart. This occurs when the spread of each Gaussian is much smaller than the distance between neighbouring Gaussian peaks $1/(s_0 \tau) \ll \Delta \phi_{\text{min}}$ where $\Delta \phi_{\text{min}}$ is the minimum gap between adjacent eigenvalues. $P(l \neq m)$ captures the overlaps between the Gaussians. This overlap contribution vanishes for large $N$, so for simplicity we will neglect this term. This neglecting will not affect the overall validity of our result. We can now write
\begin{align}\label{eq:probDelta}
P_{\Delta_{\text{E}}}>P(l=m)=\text{erf} \left(s_0 \tau \Delta_{\text{E}}\right).
\end{align}
By demanding $T_{\text{measure}} < T_{\text{bound}}$, then using Eqs. ~\eqref{eq:tprob1} and ~\eqref{eq:probDelta}, we find it is sufficient to satisfy
\begin{gather}\label{eq:Tbound}
T_{\text{bound}} \text{erf}(\tau s_0 \Delta_{\text{E}}) \gtrsim 1.
\end{gather}
For large $\tau s_0 \Delta_{\text{E}}$, the above inequality is automatically satisfied. This assumes that $\tau s_0$ grows more quickly in $N$ than the inverse of the eigenvalue uncertainty $\Delta_{\text{E}}$ that we are willing to tolerate. More generally however, it is the time and squeezing resources
we want to minimise for a given precision, so $\tau s_0 \Delta_{\text{E}}$ is small. In this case, Eq.~\eqref{eq:Tbound} becomes
\begin{gather}
T_{\text{bound}}\tau s_0 \Delta_{\text{E}} \gtrsim 1.
\end{gather}
\section{Retrieving eigenvectors in the power of one qumode}
Here we provide a brief argument of how eigenvectors of the Hamiltonian $\{ \ket{\phi_j} \}$ can also be found using our model. The hybrid state $\rho_{\text{total}}$ after application of the hybrid gate is
\begin{align}
\rho_{\text{total}} =&\frac{1}{2^n} \iint G(x) G^*(x') e^{i(x-x')H \tau/x_0} \ket{x}\!\bra{x'} \text{d}x \, \text{d}x' \nonumber \\
                             =&\sum_m \frac{1}{2^n} \iint G(x) G^*(x') e^{i(x-x')\phi_m \tau/x_0} \nonumber \\
                             & \times \ket{\phi_m}\! \bra{\phi_m} \otimes \ket{x}\!\bra{x'} \text{d}x \, \text{d}x'.
\end{align}
After a momentum measurement we are in the following state of the target register
\begin{align}
&\bra{p} \rho_{\text{total}} \ket{p} \nonumber\\
&=\frac{1}{2^n} \sum_m \mathcal{G}(\phi_m \tau/x_0-p) \mathcal{G}^*(p-\phi_m \tau/x_0) \ket{\phi_m}\! \bra{\phi_m}.
\end{align}
For a squeezed state  $G(x) = (1 / (\sqrt{s}\pi^{\frac{1}{4}})) \text{exp}(-x^2/(2s^2))$ the final state of the target register becomes
\begin{align}
\bra{p} \rho_{\text{total}} \ket{p}=\frac{s}{2^n \sqrt{\pi}} \sum_m e^{-s^2(p-\phi_m \tau/x_0)^2} \ket{\phi_m}\! \bra{\phi_m}.
\end{align}
Approximate eigenvectors can thus be obtained by measurement of the target state. The probability of obtaining the eigenvectors of the Hamiltonian is distributed in the same way as for the eigenvalues. Eigenvector identification therefore also improves with an increase in the squeezing factor. 
\section{Number of measurements for normalised trace of $U$}
Here we derive the number of momentum measurements $T_{\text{DQC1}}$ in our model needed to to recover the normalised trace of $U \equiv \exp{iH}$ to within error $\delta$. We show this is upper bounded by a quantity independent of the size of $U$. 

Let us begin by introducing a new random variable $y \equiv \exp(ip_{\text{E}} x_0)$ where $p_{\text{E}}$ are the measurement outcomes from our model. The probability distribution function with respect to $y$ can be rewritten as
\begin{gather} \label{eq:py}
\mathcal{P}_y(y)=\int^{\infty}_{-\infty} \delta(y-e^{ip_{\text{E}}x_0}) \mathcal{P}(p_{\text{E}}) \text{d}p_{\text{E}},
\end{gather}
where $\mathcal{P}(p_{\text{E}})$ is given by Eq.~\eqref{eq:pp}. 
We find that the average of $y$ is related to the normalised trace of unitary matrix $U$ 
\begin{align}
 \int y \mathcal{P}_y(y) \text{d} y &=\int e^{ip_{\text{E}}x_0} \mathcal{P}(p_{\text{E}}) \text{d}p_{\text{E}} \nonumber \\
         &=e^{-\frac{1}{4 s_0^2}} \left[\frac{\tr(U_{\tau})}{2^n}\right].
\end{align}
We now let $\tau=1$ since $U_{\tau=1}=U$. 

To find the normalised trace of $U$ to error $\delta$ is equivalent to finding the average of $y$ to within $\epsilon$ where
\begin{gather}
\int y\mathcal{P}_y(y) \text{d} y \pm \epsilon=e^{-\frac{1}{4 s_0^2}}\left(\frac{\tr(U)}{2^n} \pm \delta \right).
\end{gather}
Therefore 
\begin{gather}\label{eq:epsilondelta}
\epsilon=e^{-\frac{1}{4 s_0^2}} \delta.
\end{gather}
For concreteness, we will first separately examine recovering the real part of the normalised trace of $U$ to within $\re(\delta)$ then the imaginary part of the trace to within $\im(\delta)$.
\textit{Real part of the normalised trace of $U$.---}
We define a new random variable $y_R\equiv \re(y)=\cos(p_{\text{E}}x_0)$ whose average is within $\re(\epsilon)$ of the real part of the normalised trace of $U$. The probability distribution with repect to $y_R$ is
\begin{gather}
\mathcal{P}_{y_R}(y_R)=\int^{\infty}_{-\infty} \delta(y_R-\cos(p_{\text{E}} x_0)) \mathcal{P}(p_{\text{E}}) \text{d}p_{\text{E}}.
\end{gather}
We can employ the central limit theorem \footnote{Since we are selecting our random variable independently and from the same distribution which has finite mean and variance, it is valid to use the central limit theorem} and Eq.~\eqref{eq:epsilondelta} to find the number $t_R$ of necessary $p_{\text{E}}$ measurements to be
\begin{gather}\label{eq:clt}
t_R \sim \frac{\Sigma^2_{R}}{\re(\epsilon)^2}=\frac{\Sigma^2_{R} e^{\frac{1}{2 s_0^2}}}{\re(\delta)^2},
\end{gather}
where $\Sigma^2_R$ is the variance of the probability distribution with respect to $y_R$. Using Eqs.~\eqref{eq:py} and ~\eqref{eq:pp} we can show
\begin{align} \label{eq:SigmaR}
\Sigma^2_R \equiv & \int y^2_R \mathcal{P}_{y_R}(y_R) \text{d} y_R-\left(\int y_R \mathcal{P}_{y_R}(y_R) \text{d} y_R\right)^2 \nonumber \\
                    = &\int \cos^2(p_{\text{E}}x_0) \mathcal{P}(p_{\text{E}}) \text{d}p_{\text{E}}-\left(\int \cos(p_{\text{E}}x_0) \mathcal{P}(p_{\text{E}}) \text{d}p_{\text{E}} \right)^2 \nonumber \\
                    = &e^{-\frac{1}{2 s_0^2}}\text{sinh}\left(\frac{1}{2 s_0^2}\right) \nonumber \\
                   &+e^{-\frac{1}{4 s_0^2}}\frac{1}{2^n} \sum_{m=1}^{2^n} \cos^2(\phi_m)-e^{-\frac{1}{2 s_0^2}}\left(\frac{1}{2^n} \sum_{m=1}^{2^n} \cos(\phi_m) \right)^2 \nonumber \\
                    \leq & e^{-\frac{1}{2 s_0^2}}\text{sinh}\left(\frac{1}{2 s_0^2}\right)+e^{-\frac{1}{4 s_0^2}}\frac{1}{2^n} \sum_{m=1}^{2^n} \cos^2(\phi_m) \nonumber \\
                    \leq & e^{-\frac{1}{2 s_0^2}}\left(\text{sinh}\left(\frac{1}{2 s_0^2}\right)+e^{-\frac{1}{2s_0^2}} \right).
\end{align}
We can now use Eqs.~\eqref{eq:clt} and ~\eqref{eq:SigmaR} to find an upper bound to the number of measurements
\begin{gather}
t_R \lesssim \frac{F(s_0)}{\re(\delta)^2},
\end{gather}
where
\begin{gather}
F(s)=\text{sinh}\left(\frac{1}{2s_0^2}\right)+e^{-\frac{1}{2 s_0^2}}.
\end{gather}
\textit{Imaginary part of the normalised trace of $U$.---}
To recover the imaginary part of the normalised trace of $U$ to within an error $\im(\delta)$, we average $y_I \equiv \im(y)=\sin(p_{\text{E}}x_0)$. The probability distribution with respect to $y_I$ is
\begin{gather}
\mathcal{P}_{y_I}(y_I)=\int^{\infty}_{-\infty} \delta(y_I-\sin(p_{\text{E}}x_0)) \mathcal{P}(p_{\text{E}}) \text{d}p_{\text{E}}.
\end{gather}
We can similarly use the central limit theorem in this case to find the necessary number of measurements $t_I$ 
\begin{gather}
t_I \sim \frac{\Sigma^2_{I}e^{\frac{1}{2 s_0^2}}}{\im(\delta)^2},
\end{gather}
where $\Sigma^2_I$ is the variance with respect to probability distribution $P_{y_I}(y_I)$. We can show
\begin{align}
\Sigma^2_I \equiv & \int y^2_I \mathcal{P}_{y_I}(y_I) \text{d} y_I-\left(\int y_I \mathcal{P}_{y_I}(y_I) \text{d} y_I\right)^2 \nonumber \\
               =& e^{-\frac{1}{2 s_0^2}}\text{sinh}\left(\frac{1}{2 s_0^2}\right) \nonumber \\
                   &+e^{-\frac{1}{4 s_0^2}}\frac{1}{2^n} \sum_{m=1}^{2^n} \sin^2(\phi_m)-e^{-\frac{1}{2 s_0^2}}\left(\frac{1}{2^n} \sum_{m=1}^{2^n} \sin(\phi_m) \right)^2 \nonumber \\
               \leq & e^{-\frac{1}{2 s_0^2}} F(s_0).
\end{align}
Thus
\begin{gather}
t_I \lesssim \frac{F(s_0)}{\im(\delta)^2}.
\end{gather}
This means the number of required measurements $t$ to recover the normalised trace of $U$ to within $\delta$ has the upper bound
\begin{gather}
T_{\text{DQC1}} =\text{max}(t_R,t_I) \lesssim \frac{F(s_0)}{[\text{min}\{ \text{Re}(\delta), \text{Im}(\delta)\}]^2}.
\end{gather}
\section{Number of measurements needed for factoring in the power of one qumode}
Here we give the derivation of the number of runs $T_{\text{factor}}$ needed to recover a non-trivial factor of $N$ given the momentum probability distribution (Eq.~\eqref{eq:PDFfactor})
\begin{gather}
\mathcal{P}(p_{\text{E}})=\frac{s_0 \tau}{\sqrt{\pi} 2^n} \sum_{m=0}^{r-1} c_m e^{-(2 \pi s_0 \tau)^2 \left(\frac{p_{\text{E}}}{2 \pi}-\frac{m}{r}\right)^2}.
\end{gather}
We want to find the probability $P_{r}$ in which one can retrieve the correct value of the order $r$. The number of runs required on average to find a non-trivial factor of $N$ is inversely related to this probability
\begin{gather} \label{eq:timefactor}
T_{\text{factor}} \sim \frac{1}{P_{r}}.
\end{gather}
Here we derive a lower bound to $P_{r}$ (hence an upper bound to the number of runs) that satisfies the following two conditions. To recover $r$ it is sufficient to (i) know $m/r$ to an accuracy within $1/(2N^2)$ and (ii) to choose when $m$ and $r$ have no factors in common so their greatest common denominator is one (i.e., $\gcd(m,r)=1$). 

The first condition comes from the continued fractions algorithm \cite{nandc}, which can be used to exactly recover the rational number $m/r$ given some $\phi'$ when $\left|\phi'-m/r\right|\leq 1/(2r^2)$. Since $r \leq N$, a sufficient condition is $\left|\phi'-m/r\right|\leq 1/(2N^2)$. The second condition ensures we recover $r$ instead of a non-trivial factor of $r$. We will see how to satisfy the second condition later on.

To satisfy the first condition, we see that the probability of finding $m/r$ to within $1/(2 N^2)$ when measuring $p'_{\text{E}} \equiv p_{\text{E}}/(2 \pi)$ is
\begin{align}
P_{r}&\equiv \mathcal{P} \left(p'_{\text{E}};\left|p'_{\text{E}}-\frac{m}{r}\right| \leq \frac{1}{2N^2}\right) \nonumber \\
                   & =\frac{s_0 \tau}{\sqrt{\pi}2^n} \sum_{l=0}^{r-1} \int^{\frac{l}{r}+\frac{1}{2N^2}}_{\frac{l}{r}-\frac{1}{2N^2}} \sum_{m=0}^{r-1} c_m e^{-(2 \pi s_0 \tau)^2 \left(p'_{\text{E}}-\frac{m}{r}\right)^2} 2\pi \, \text{d}p'_{\text{E}} \nonumber \\
                  &>\frac{s_0 \tau}{\sqrt{\pi}2^n} \sum_{m=0}^{r-1} c_m \int^{\frac{m}{r}+\frac{\pi}{2N^2}}_{\frac{m}{r}-\frac{\pi}{N^2}} e^{-(2 \pi s_0 \tau)^2 (p'_{\text{E}}-\frac{m}{r})^2} 2 \pi \text{d}p'_{\text{E}} \nonumber \\
             &= \sum_{m=0}^{r-1} \frac{c_m}{2^n} \text{erf} \left(\frac{\pi s_0 \tau}{N^2}\right)=\sum_{m=0}^{r-1} \frac{c_m}{2^n} \text{erf} \left(\frac{\pi s_0 \tau}{2^{2n}}\right).
\end{align}
Note that we do not require contributions to the probability from every $m$ in the summation. In order to successfully retrieve $r$ from the fraction $m/r$, we
need only consider the cases where $\gcd (m,r)=1$. Euler's totient function $\Phi(r)$ represents the number of cases where $m$ and $r$ are coprime with $m<r$. It can be
shown that $\Phi(r) >r/\{e^{\gamma} \ln[\ln(r)]\}$ where $\gamma$ is Euler's number \cite{shor}. In the cases where $\gcd (m,r)=1$, the amplitude
$|c_m| \equiv M$, where $M$ is the number of cases where $r_d=r$. It is also possible to show that when $N=v_1v_2$ (where $v_1$ and $v_2$ are prime numbers),
$M> (v_1-1)(v_2-1)$ \cite{parkerplenio}. 

Then the probability of retrieving the correct $r$ from the probability distribution is at least
\begin{align}\label{eq:prbound}
P_{r} &>\sum_{m=0}^{r-1} \frac{c_m}{2^n} \text{erf} \left(\frac{\pi s_0 \tau}{2^{2n}}\right)>\frac{M \Phi(r)}{2^n} \text{erf}\left(\frac{\pi s}{2^{2n}}\right) \nonumber \\
                    &> \frac{(v_1-1)(v_2-1) r}{e^{\gamma} 2^n \ln[\ln(r)]} \text{erf}\left(\frac{\pi s_0 \tau}{2^{2n}}\right) \nonumber \\
                    &>  \frac{(v_1-1)(v_2-1)}{e^{\gamma} 2^{n} \ln[\ln(r)]} \text{erf}\left(\frac{\pi s_0 \tau}{2^{2n}}\right).
\end{align}
From Eqs.~\eqref{eq:timefactor} and ~\eqref{eq:prbound} we now have an upper bound to the time steps required
\begin{gather}\label{eq:tfactorize}
T_{\text{factor}} <\frac{1}{P_r}=\frac{e^{\gamma} N \ln[\ln(N)]}{(v_1-1)(v_2-1) \text{erf}\left(\frac{\pi s_0 \tau}{2^{2n}}\right)}.
\end{gather}
The large $N$ limit (where $v_1,v_2 \gg 1$) gives our result 
\begin{gather}
T_{\text{factor}}<\frac{e^{\gamma}  \ln[\ln(N)]}{\text{erf}\left(\frac{\pi s_0 \tau}{2^{2n}}\right)}=\frac{e^{\gamma}  \ln[\ln(2^n)]}{\text{erf}\left(\frac{\pi s_0 \tau}{2^{2n}}\right)}.
\end{gather}
\end{document}